%% file: EduCHI_ArXiv.tex
\documentclass[sigconf]{acmart}

\usepackage{graphicx}
\usepackage{xcolor}
\usepackage{booktabs}
\usepackage{graphicx}

\usepackage{pgfplots}
\pgfplotsset{compat=1.18}
\usepgfplotslibrary{groupplots}

\usepackage[most]{tcolorbox}

\usepackage{tikz}
\usetikzlibrary{arrows.meta, positioning, shapes, fit, calc}

\AtBeginDocument{%
  }

\copyrightyear{2026}
\acmYear{2026}
\setcopyright{cc}
\setcctype{by}
\acmConference[EduCHI '26]{EduCHI 2026: 8th Annual Symposium on HCI Education}{May 20--22, 2026}{Toronto, ON, Canada}
\acmBooktitle{EduCHI 2026: 8th Annual Symposium on HCI Education (EduCHI '26), May 20--22, 2026, Toronto, ON, Canada}
\acmDOI{10.1145/3803869.3803883}
\acmISBN{979-8-4007-2430-5/2026/05}

\begin{document}

\title[Teaching Usable Privacy in HCI Education]{Teaching Usable Privacy in HCI Education: Designing, Implementing, and Evaluating an Active Learning Graduate Course}

\author{Sanchari Das}
\email{sdas35@gmu.edu}
\affiliation{%
  \institution{George Mason University}
  \city{Fairfax}
  \state{Virginia}
  \country{USA}
}

\author{Dhiman Goswami}
\email{dgoswam@gmu.edu}
\affiliation{%
  \institution{George Mason University}
  \city{Fairfax}
  \state{Virginia}
  \country{USA}
}

\author{Michelle Melo}
\email{mmelo4@gmu.edu}
\affiliation{%
  \institution{George Mason University}
  \city{Fairfax}
  \state{Virginia}
  \country{USA}
}

\author{Aditya Johri}
\email{johri@gmu.edu}
\affiliation{%
  \institution{George Mason University}
  \city{Fairfax}
  \state{Virginia}
  \country{USA}
}

\author{Vivian G. Motti}
\email{vmotti@gmu.edu}
\affiliation{%
  \institution{George Mason University}
  \city{Fairfax}
  \state{Virginia}
  \country{USA}
}

\renewcommand{\shortauthors}{Das et al.}

\begin{abstract}
As digital systems increasingly rely on pervasive data collection and inference, educating future designers and researchers about~\textit{Usable Privacy} has become a critical need for HCI. However, privacy education in higher education is often fragmented, theory-heavy, or detached from real-world applications. Thus, in this paper, we present the design, implementation, and evaluation of a 15-week graduate-level course on~\textit{Usable Privacy} that addresses this through active, practice-oriented pedagogy. The course integrates use cases, structured role playing, case-based discussions, guest lectures, and a multi-phase research project to support students in reasoning about privacy from multiple stakeholder perspectives. Grounded in contemporary privacy research and the \emph{Modern Privacy} framework, the curriculum emphasizes both conceptual understanding and applied research skills. We report findings from two course offerings in consecutive years (2024-2025) using a mixed-methods evaluation that combines quantitative teaching evaluations with qualitative analysis of student reflections and instructor observations. Results indicate increased student engagement, improved ability to articulate trade-offs in privacy design, and stronger connections between theory and practice. To support adoption and replication, we also release detailed assignment descriptions and grading rubrics. This work contributes an empirically informed model for teaching~\textit{Usable Privacy} in HCI education and offers actionable guidance for educators seeking to integrate privacy into their curricula.
\end{abstract}

\ccsdesc[500]{Human-centered computing~Human computer interaction (HCI)}
\ccsdesc[500]{Human-centered computing~Empirical studies in HCI}
\ccsdesc[500]{Human-centered computing}
\ccsdesc[500]{Security and privacy~Usability in security and privacy}

\keywords{Usable Privacy, Higher Education, Graduate Course, Education Intervention, Use Cases, Role Playing}

\maketitle

\section{Introduction and Motivation}
Digital systems operate within an ecosystem of pervasive data collection, large-scale aggregation, and automated inference. Mobile devices, wearables, smart homes, and AI-driven platforms continuously generate behavioral traces that organizations integrate across contexts and services~\cite{xi2025integrating,ahmed2022overview}. Advances in storage, computation, and machine learning further enable inference of sensitive attributes users neither disclose nor anticipate~\cite{liang2018deep,shokri2015privacy,shrestha2022exploring}. As a result, privacy risks extend beyond data collection to secondary use, cross-platform integration, profiling, and opaque algorithmic processing~\cite{sun2024privacy,bentotahewa2024profiling}, exacerbating asymmetries of power, visibility, and control between users and institutions~\cite{coll2014power,gilman2018surveillance,adhikari2022privacy}. Questions of data ownership and governance intensify these tensions.

In response, Usable Privacy has matured as an interdisciplinary field spanning HCI, security, behavioral science, law, and socio-technical systems research~\cite{de2025user,garfinkel2014usable,hutt2023right,das2019security,leutenegger2026starting}. Research shows that formal guarantees and regulatory safeguards fail when systems misalign with users' mental models, cognitive limits, contextual expectations, and lived experiences~\cite{windl2025human,stojkovski2022user,ganguly2026aligning}. Scholars document breakdowns in consent interfaces~\cite{bouma2023us}, problematic defaults~\cite{pilavakis2023didn}, and usability barriers in privacy-enhancing technologies (PETs)~\cite{klymenko2025we}, underscoring that privacy must be addressed as a socio-technical design challenge. Yet privacy education remains fragmented, often siloed within law or technical security courses, limiting integrated socio-technical reasoning~\cite{swart2007framework,denardis2020researching}. Many usable security/privacy courses emphasize paper critique or isolated projects without coherent scaffolding across theory, stakeholders, and empirical methods, and static approaches risk rapid obsolescence. Preparing students for contemporary privacy challenges requires integrating conceptual foundations, stakeholder perspective-taking, applied research skills, and iterative reflection. Students must reason through trade-offs among usability, security, accessibility, business incentives, and regulation~\cite{naqvi2019interdependencies,distler2020acceptable,wijenayake2025advancing}. Situated and active learning theories suggest that engagement with authentic problems and negotiated perspectives deepens understanding~\cite{wilson2024constructivism,mclellan1996situated}, particularly in domains like privacy where ethical tensions resist simple solutions~\cite{ijaiya2024balancing,pozen2016privacy}.

To address this, we designed and implemented \textit{Usable Privacy in the Digital Age}, a 15-week graduate course grounded in situated and active learning. We integrated role play, case-based debate, forum discussions, guest lectures, and a scaffolded multi-phase research project to position privacy as a dynamic socio-technical phenomenon. The curriculum connects foundational theory with domains such as social media, IoT, healthcare, accessibility, and AI-mediated systems. We evaluate two consecutive offerings using mixed methods, combining quantitative teaching evaluations with qualitative reflections to assess engagement and trade-off reasoning. 

\paragraph{Contributions.}
In this paper, we contribute a reusable, practice-oriented approach for teaching usable privacy in HCI programs. Specifically, we:

\begin{itemize}
    \item Design and articulate a three-phase instructional structure that integrates socio-technical foundations with active learning strategies, including role play, case-based discussion, guest lectures, and guided research practice.
    
    \item Present a detailed implementation blueprint for a 15-week graduate course, including aligned assessments and scaffolding mechanisms that instructors can adapt to diverse institutional contexts.
    
    \item Provide mixed-method evaluation evidence from two consecutive course offerings, combining standardized teaching evaluations and qualitative reflections to document student engagement and perceived learning outcomes.
    
    \item Share transferable teaching artifacts and activity patterns (e.g., role-play templates, case-study workflows, and a multi-phase project sequence) to support replication, adaptation, and iterative refinement by HCI educators.
\end{itemize}

\section{Related Work}

\subsection{Theoretical Framework}
While privacy and usable privacy have long been taught, our objective was to cultivate contextual and socio-technical reasoning. We grounded the course in situated learning theory~\cite{greeno1996cognition,greeno1998situativity,greeno2011situative,johri2014situative,lave1991situated}, which conceptualizes knowledge as dynamically constructed through participation in context rather than stored as abstract content. From this perspective, learning depends on how individuals engage with situations, interpret information, and negotiate meaning. This theoretical grounding directly informed our use of case studies and role plays, which situate privacy reasoning within realistic socio-technical contexts.

\paragraph{Active Learning}
Traditional lecture-dominant instruction often limits student engagement and relies on low-level participation \cite{mccarthy2000active,grabinger1996active,prince2004active}. Although historically justified by assumptions about limited prior knowledge \cite{grabinger1996active}, research consistently demonstrates that active learning—collaborative learning and problem-based learning (PBL)—improves engagement and outcomes \cite{prince2004active}. Longitudinal interventions show increased motivation \cite{julia2019impact}, stronger engagement in real-world problem solving \cite{julia2019impact}, improved inquiry skills in STEM contexts \cite{educsci12100686}, and graduate student preference for active strategies \cite{patrick2021roles}. Active learning also strengthens instructor–student interaction and peer collaboration \cite{capone2022blended}. These findings justify structuring the course around participatory, discussion-driven formats.

\paragraph{Case Study and Guest Lectures}
Case-based teaching, originating in business education \cite{herzog2025cases}, centers on narrative scenarios requiring decision-making and contextual analysis \cite{herzog2025cases}. Effective implementation demands careful design and active student preparation \cite{hackney2003using}, and may take lecture, discussion, small-group, or individual formats \cite{herreid2011case}. Case methods promote higher-order reasoning, experiential learning, integration of multiple perspectives \cite{hackney2003using}, and respect for diverse viewpoints \cite{herreid2011case}. Complementing this approach, guest lectures enhance learning by addressing diverse needs \cite{alshahrani2024use,pepple2025using} and strengthening connections between theory and professional practice \cite{10.18538/lthe.v13.n2.229}. Together, these pedagogies situate privacy within authentic institutional and societal contexts.

\paragraph{Role Playing}
Role play extends case-based learning through simulation \cite{hertel2023role-play}, ensuring the articulation of stakeholders perspectives. It supports perspective-taking, critical thinking, and communication skills \cite{hingle2024role}. Research shows that involving students in structuring role-play components enhances learning effectiveness \cite{cherif1995maximizing}, and empirical evidence indicates higher learning gains and exam performance compared to lecture-only instruction \cite{raymond2010role}. These findings justify our use of structured stakeholder role play to operationalize socio-technical trade-off reasoning in privacy education.

\subsection{Usable Privacy}
Usable Privacy lies at the intersection of HCI, privacy, security, and socio-technical systems research. It examines how individuals understand, experience, and act upon privacy in interactive systems, and how designers can create mechanisms that both protect users and remain usable in practice~\cite{jacobs2022survey}. Unlike purely technical security approaches that prioritize formal guarantees, usable privacy foregrounds human factors such as mental models, risk perception, cognitive load, and contextual decision-making~\cite{coopamootoo2014mental}. Early research identified a persistent gap between formal privacy controls and actual user behavior, showing that users struggle to interpret notices, configure settings, and exercise meaningful consent in complex interface environments~\cite{schraefel2017internet}. Privacy decision-making is deeply shaped by contextual cues and interface design, and usability breakdowns can undermine well-intentioned protections~\cite{alemany2022review}. Empirical studies further demonstrate misalignment between privacy interfaces and users' expectations, leading to misconfiguration, habituation, and disengagement~\cite{lin2012expectation,o2025cybersecurity}.

A second research stream examines behavioral and economic dimensions of privacy. The ``privacy paradox" illustrates divergence between stated preferences and actual behavior~\cite{kokolakis2017privacy}. Framing effects, defaults, and nudges significantly influence disclosure decisions~\cite{acquisti2017nudges}, underscoring the need to account for bounded rationality and cognitive bias in design~\cite{waldman2020cognitive}. Work on PETs similarly shows that technical soundness alone is insufficient; adoption depends on usable interfaces and transparent communication~\cite{tavani2001privacy}. Users must be able to understand privacy–utility trade-offs for protections to be effective~\cite{calvi2024unfair}, motivating calls to integrate user-centered design principles into PET development and evaluation~\cite{feth2017user}. Usable privacy research has also expanded across domains including social media, smart homes, healthcare, IoT, and child-directed platforms~\cite{zheng2018user,chhetri2022user,Chhetri2020IdentifyingVI,zhang2023study}. These contexts introduce stakeholder tensions, secondary use, inference risks, and power asymmetries that challenge consent-centric models. Privacy expectations shift in shared and ambient environments where data collection is continuous and partially invisible~\cite{langheinrich2002privacy}, reinforcing the importance of contextual integrity and negotiated boundaries~\cite{nissenbaum2004privacy}.Recent scholarship frames privacy as equity-centered, socio-technical, documenting disproportionate harms experienced by marginalized populations~\cite{mcdonald2020privacy} and emphasizing structural power, accessibility, and cross-cultural norms in privacy design~\cite{ur2013cross}. This aligns with broader HCI calls to embed ethics, justice into system evaluation and design~\cite{abascal2005moving}.

Despite this rich body of work, privacy education remains fragmented. Many programs isolate privacy within security engineering or policy discussions, without integrating empirical methods, socio-technical analysis, and user-centered design into a coherent curriculum~\cite{saltarella2024translating}. There is limited published work describing structured, empirically evaluated socio-technical privacy courses that incorporate active learning, role play, and scaffolded research projects~\cite{hingle2024framework}. We respond to this gap by synthesizing empirical usable privacy research, behavioral theory, privacy engineering, and socio-technical frameworks into a structured graduate curriculum. By combining established scholarship with active pedagogy, we seek to bridge research and educational practice in usable privacy~\cite{mulligan2011bridging}.

\section{Prior Usable Privacy/Security Courses}
Dedicated courses in usable privacy and usable security have emerged across universities, but they vary substantially in scope (privacy-only vs.\ combined security and privacy), pedagogy (seminar critique vs.\ project-based research), and audience (undergraduate, graduate, professional/online). We summarize representative offerings that foreground human factors, usability, and socio-technical perspectives, and situate our course relative to them~\cite{roldan2021pedagogical,francisco2025computer,johri2024case,khadka2026sok,tazi2021parents}.

\subsection{Established Research-Oriented University Courses}
Carnegie Mellon University's \textit{Usable Privacy and Security}, offered through CyLab at both undergraduate and graduate levels, exemplifies a research-intensive model. The course combines lectures and projects, emphasizing usability challenges in security and privacy and training students to design and evaluate empirical studies in this space~\cite{cmu_usable_privacy_security}. This approach reflects a research-preparation pathway in which students critically engage with current literature while producing work aligned with publishable outcomes~\cite{pownall2023teaching}. Similarly, the University of Chicago's \textit{Usable Security and Privacy (CMSC 23210 / CMSC 33210)} bridges computer systems, HCI, and public policy, covering topics such as authentication, web security, privacy notices, warnings, and IoT contexts, typically culminating in a capstone research project~\cite{uchicago_usable_sec_spring22}. Duke University's \textit{CPS 586: Usable Security \& Privacy} foregrounds engagement with contemporary research and human-centered study design through a substantial group research project~\cite{duke_compsci586_spring26}. Tufts University's \textit{CS 152-1: Human Factors in Security and Privacy} further emphasizes research design, experimental and qualitative methods, and ethics in human-subject studies~\cite{tufts_comp152hfs}. Collectively, these courses demonstrate a common pattern: usable privacy/security is often taught through paper critique, formal methods training, and extended research projects within technically oriented computing curricula~\cite{egelman2016teaching}.

\subsection{Information Schools and Interdisciplinary Offerings}
Usable privacy/security also appears in information schools and interdisciplinary programs, frequently integrating user-centered design with foundational security concepts. The University of Wisconsin–Madison's \textit{LIS 640: Usable Security and Privacy} focuses on designing and evaluating security and privacy systems from a user-centered perspective, covering authentication, phishing, warnings, privacy notices, and IoT contexts~\cite{wisc_usable_security}. UC Berkeley's MICS curriculum includes \textit{Usable Privacy and Security Research}, positioning human factors as central to challenges such as patching behavior, password practices, phishing susceptibility, and encryption adoption, with emphasis on evaluation and design of usable mechanisms~\cite{berkeley_cyber_215}. These offerings reflect growing recognition of usable privacy/security as an interdisciplinary competency spanning HCI, policy, and applied cybersecurity practice~\cite{albarrak2024integration,tazi2023cybersecurity}.

\subsection{Special Topics and Regionally Distributed Courses}
Beyond major research institutions, usable privacy/security courses appear as special-topics offerings. The George Washington University's \textit{CSCI 4533/6533: Introduction to Usable Security and Privacy} emphasizes that provable security is insufficient without usability, outlining objectives that include paper critique, human factors methods, hypothesis formation, and study design~\cite{aviv_usec_f25}. New Jersey Institute of Technology's \textit{CS 485-007: Selected Topics in CS -- Usable Security and Privacy} further demonstrates diffusion of the field into departmental electives~\cite{njit_cs485_usable_security_privacy}. Karlstad University's \textit{Usable Security and Privacy (DVAE25)} integrates lectures, seminars, and practical assignments on legal principles, PETs, and HCI methods, including prototyping and usability evaluation~\cite{karlstad_dvae25_usable_security_privacy}. The University of Waterloo's \textit{Usable Security and Privacy (ECE 750 T38)} emphasizes stakeholder reasoning and execution of empirical studies~\cite{uwaterloo_usable_security_privacy}, while the University of Central Florida's \textit{IDC 6602: Usable Cybersecurity \& Privacy} incorporates presentations and a term project centered on usability testing and redesign~\cite{ucf_idc6602_23spring}.

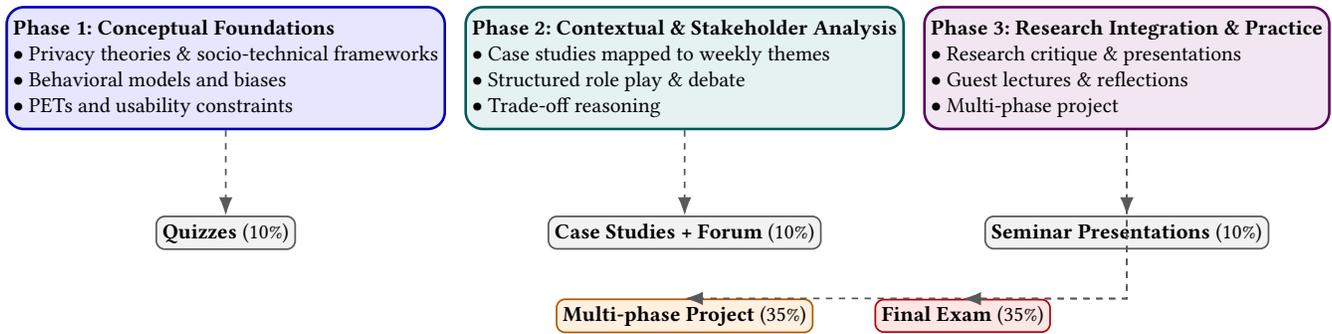
\begin{figure*}[t]
\centering
\resizebox{0.99\linewidth}{!}{%
\begin{tikzpicture}[
font=\large,
  node distance=3mm and 3mm,
  >=Latex,
  box/.style 2 args={
    rectangle,
    rounded corners=8pt,
    draw=#1,
    fill=#2,
    very thick,
    align=left,
    inner sep=3pt,
    minimum width=5.25cm,
    minimum height=20mm
  },
  tag/.style 2 args={
    rectangle,
    rounded corners=4pt,
    draw=#1,
    fill=#2,
    thick,
    inner sep=3pt,
  },
  flow/.style={-{Latex[length=3mm]}, very thick, draw=blue!70!black},
  loop/.style={-{Latex[length=3mm]}, very thick, draw=blue!70!black},
  link/.style={-{Latex[length=3mm]}, thick, dashed, draw=gray!70!black}
]

\node[box={blue!70!black}{blue!10}] (p1) {\textbf{Phase 1: Conceptual Foundations}\\
\(\bullet\) Privacy theories \& socio-technical frameworks\\
\(\bullet\) Behavioral models and biases\\
\(\bullet\) PETs and usability constraints};

\node[box={teal!70!black}{teal!10}, right=of p1] (p2) {\textbf{Phase 2: Contextual \& Stakeholder Analysis}\\
\(\bullet\) Case studies mapped to weekly themes\\
\(\bullet\) Structured role play \& debate\\
\(\bullet\) Trade-off reasoning};

\node[box={violet!70!black}{violet!10}, right=of p2] (p3) {\textbf{Phase 3: Research Integration \& Practice}\\
\(\bullet\) Research critique \& presentations\\
\(\bullet\) Guest lectures \& reflections\\
\(\bullet\) Multi-phase project};

\coordinate (laneL) at ($(p1.north west)+(0,18mm)$);
\coordinate (laneR) at ($(p3.north east)+(0,18mm)$);
\coordinate (laneC1) at ($(p1.north)+(0,18mm)$);
\coordinate (laneC2) at ($(p2.north)+(0,18mm)$);
\coordinate (laneC3) at ($(p3.north)+(0,18mm)$);

\node[tag={gray!70!black}{gray!10}, below=14mm of p1] (a1) {\textbf{Quizzes} (10\%)};
\node[tag={gray!70!black}{gray!10}, below=14mm of p2] (a2) {\textbf{Case Studies + Forum} (10\%)};
\node[tag={gray!70!black}{gray!10}, below=14mm of p3] (a3) {\textbf{Seminar Presentations} (10\%)};

\node[tag={orange!70!black}{orange!12}, below=8mm of a2] (a4) {\textbf{Multi-phase Project} (35\%)};
\node[tag={red!70!black}{red!10}, right=10mm of a4] (a5) {\textbf{Final Exam} (35\%)};

\draw[link] (p1.south) -- (a1.north);
\draw[link] (p2.south) -- (a2.north);
\draw[link] (p3.south) -- (a3.north);

\draw[link] (p3.south) |- (a4.north);
\draw[link] (p3.south) |- (a5.north);

\end{tikzpicture}%
}
\caption{Three-phase instructional progression with iterative reinforcement and aligned assessments.}
\Description{A three-phase course design diagram arranged left to right. 
Phase 1 (Conceptual Foundations) includes privacy theories, socio-technical frameworks, behavioral models, and privacy-enhancing technologies. 
Phase 2 (Contextual and Stakeholder Analysis) includes case studies, structured role play, and trade-off reasoning. 
Phase 3 (Research Integration and Practice) includes research critique, guest lectures, and a multi-phase project. 
Arrows indicate forward progression and iterative reinforcement across phases. 
Below each phase are aligned assessments: quizzes (25\%), case and forum participation (20\%), presentations (15\%), a multi-phase project (35\%), and participation(5\%).}
\label{fig:course_flow}
\end{figure*}

\subsection{Synthesis and Gap Relative to HCI Privacy Education}
Across these offerings, a clear pattern emerges. Courses frequently combine usable security and privacy, prioritize research-method training, and center seminar-style literature critique or capstone research projects~\cite{uchicago_usable_sec_spring22,duke_compsci586_spring26,tufts_comp152hfs,egelman2016teaching}. While these models provide strong foundations, fewer published curricula articulate a full-semester, privacy-centered sequence tailored specifically to HCI programs that integrates socio-technical theory, structured stakeholder role play, guided case deliberation, guest lectures with reading reflections, and a scaffolded multi-phase research project. Our course complements and extends prior work by positioning privacy as a first-class HCI design and research concern while packaging reusable assignments and rubrics to support adoption in HCI-focused contexts~\cite{benbenisty2021privacy}.

\section{\textit{Usable Privacy in the Digital Age}: Course Curriculum}

\textit{Usable Privacy in the Digital Age} is a 15-week graduate-level course intentionally structured to integrate socio-technical privacy theory, empirical research methods, and applied design reasoning within a cohesive pedagogical framework. Delivered in person, the course guides students through a progressive learning trajectory that builds conceptual foundations, cultivates stakeholder-centered analysis, and strengthens research competence. Rather than presenting privacy as a fixed technical safeguard or compliance requirement, the curriculum positions it as a dynamic socio-technical challenge shaped by cognitive processes, power structures, institutional incentives, regulatory environments, and evolving technological infrastructures. The assessment structure reinforces this integrated approach. Students engage in formative and summative evaluations including quizzes ($10\%$), research presentations ($10\%$), structured case-based engagement and forum discussions ($10\%$), a scaffolded multi-phase research project ($35\%$), and final examination ($35\%$). This distribution deliberately balances foundational knowledge acquisition, applied critical reasoning, and sustained empirical inquiry, ensuring that students demonstrate both theoretical mastery and the ability to operationalize privacy principles in real-world contexts. Figure~\ref{fig:course_flow} presents the pedagogical structure of the course, highlighting the sequential progression across phases and the iterative reinforcement of theory through applied practice. There are no prerequisites for enrollment to ensure that students from diverse disciplinary backgrounds can fully participate in the course’s socio‑technical approach to privacy.

\subsection{Foundational Text and Conceptual Anchoring}
The course is anchored in \textit{Modern Socio-Technical Perspectives on Privacy}~\cite{knijnenburg2022modern}, an interdisciplinary volume that synthesizes theoretical, behavioral, technical, and design-oriented approaches to privacy. The text frames privacy as a socio-technical construct shaped by contextual norms, individual cognition, institutional incentives, and technological affordances, positioning it not as a static legal right or purely technical property, but as an evolving phenomenon emerging from interactions among users, systems, and governance structures. The volume introduces foundational theories such as contextual integrity and socio-cognitive models of decision-making, alongside behavioral perspectives including the privacy paradox and bounded rationality. It examines how mental models, heuristics, and framing effects shape disclosure behavior, and surveys PETs with attention to both technical properties and usability constraints. It also addresses cross-cultural norms, vulnerable populations, and ethical design, linking privacy to accessibility, power, and equity. In the course, the book functions as a conceptual backbone rather than a linear reading sequence. Early weeks establish shared vocabulary and analytical frameworks, which are then revisited and operationalized throughout the semester. Behavioral models inform discussions of consent and defaults; contextual integrity frames analyses of social media data flows; and PET frameworks ground evaluations of IoT and AI-mediated systems. This recursive integration maintains theoretical coherence while extending to contemporary domains such as personalization, healthcare ecosystems, smart environments, and algorithmic inference. The text thus serves as a durable scaffold supporting cumulative reasoning, critical analysis, and methodological rigor across the course.

\subsection{Instructional Design and Learning Progression}
The curriculum is based on three interrelated phases.
\vspace{-2mm}
\paragraph{Phase 1: Conceptual Foundations.}
The course begins by situating privacy historically and philosophically before introducing contemporary models and behavioral perspectives. Lectures examine privacy decision-making, cognitive biases, tracking infrastructures, personalization systems, inference risks, and privacy-enhancing technologies. Ethical considerations, accessibility, and privacy for marginalized populations are embedded throughout. Weekly quizzes reinforce comprehension, while short in-class exercises prompt immediate engagement with emerging scenarios.

\paragraph{Phase 2: Stakeholder-Centered Case Analysis.}
Following major thematic units, students engage weekly in structured case studies that require adopting stakeholder roles such as platform designer, policymaker, user, or an affected community member. These moderated debates cultivate perspective-taking and trade-off reasoning. For example, during the analysis of the Equifax data breach~\cite{zou2018ve}, students examined systemic vulnerabilities and conflicting incentives across institutional and individual actors. Prior to in-class discussions and case analysis, students submit structured position statements to an online forum, encouraging preparation, accountability, and asynchronous reflection. 

\begin{tcolorbox}[
colback=gray!6,
colframe=black!60,
title=\textbf{Case Study A: Fictionalized Identity Theft Scenario},
fonttitle=\normalsize,
enhanced,
breakable
]

Alice's nightmare with identity theft began in 2012 when she received a text message stating that her mobile number was being ported to another network provider without her authorization. Shortly thereafter, she lost access to her banking account and began receiving notifications of unauthorized changes to her contact information and credit card credentials. Despite contacting her mobile provider and financial institutions, she was unable to regain control before the attacker reset multiple account credentials using her compromised phone number.

Even after requesting account locks, fraudulent credit activity continued. New credit cards she did not apply for arrived at her residence. Over the following months, Alice spent countless hours contacting companies, navigating long support queues, and attempting to prove that she was not responsible for the fraudulent activity. Her credit score, built over years of effort, collapsed within months.

Years later, upon learning of the 2017 Equifax breach, she checked whether her information had been exposed. Seeing the notice that her personal data may have been impacted reignited fear, frustration, and emotional distress. The ongoing uncertainty reinforced a sense that identity theft was an endless, recurring threat.

\textbf{Source:} Chen, J. X., McDonald, A., Zou, Y., Tseng, E., Roundy, K. A., Tamersoy, A., Schaub, F., Ristenpart, T., \& Dell, N. (2022). \textit{Trauma-informed computing: Towards safer technology experiences for all}. Proceedings of the 2022 CHI Conference on Human Factors in Computing Systems (pp. 1–20). ACM. \url{https://doi.org/10.1145/3491102.3502007}~\cite{chen2022trauma}

\textbf{Stakeholder Roles}
\begin{itemize}
    \item Identity theft victim
    \item Law enforcement and financial institutions
    \item Scammer / identity thief
    \item Mobile service provider
    \item Credit reporting agencies
    \item Customer support representatives
    \item Regulators and policymakers
\end{itemize}

\textbf{Objective}
What actions would your assigned stakeholder take? What responsibilities should be upheld, and what responses should be avoided?

\textbf{Assessment Structure}
\begin{itemize}
    \item Online discussion board (50\%): Written position prior to class.
    \item In-class discussion (50\%): Live role-based debate and response to peer challenges.
\end{itemize}

\end{tcolorbox}

This case study presents a fictionalized identity-theft scenario used in the course to support structured stakeholder role-play and socio-technical analysis. The narrative highlights cascading failures across telecommunications, banking, and credit infrastructures, enabling students to examine breakdowns in usability, accountability, and institutional response. By foregrounding both technical vulnerabilities and the emotional impact on the victim, the case encourages discussion of trauma-informed privacy design and the broader societal implications of data misuse.

\paragraph{Phase 3: Research Integration and Practice.}
The final phase of the course curriculum emphasizes research literacy and applied inquiry. For that, students analyze contemporary usable privacy scholarship through group presentations that critically evaluate research questions, methodologies, and contributions. Structured reading reflections prepare students for guest lectures, fostering deeper engagement with active researchers and current debates in the field. The multi-phase final project forms the central integrative experience of the course. Working in small groups, students first conduct a structured literature review on a chosen usable privacy topic, then perform an analytical or experimental evaluation that examines privacy–utility trade-offs or system design implications, and finally design and execute a pilot user study informed by their earlier findings. This progression mirrors the lifecycle of empirical research and requires students to synthesize theory, technical analysis, and human-centered evaluation. Rather than isolated assignments, these phases build cumulatively, reinforcing methodological rigor and contextual reasoning.

The course concludes with a cumulative final examination designed to assess students' conceptual mastery and their ability to synthesize ideas across the full spectrum of topics covered throughout the semester. Rather than focusing solely on factual recall, the examination evaluates students' capacity to integrate socio-technical theory, behavioral models, privacy-enhancing technologies, and stakeholder-based reasoning into coherent analytical responses. Questions require students to articulate trade-offs, apply foundational frameworks to novel scenarios, and demonstrate cross-topic connections between theory, case analysis, and research practice. In this way, the final assessment reinforces the course's emphasis on durable reasoning skills and holistic understanding rather than isolated knowledge acquisition.

\subsection{Pedagogical Coherence}
Across lectures, stakeholder debates, research activities, and assessments, the curriculum maintains intentional alignment between theory, context, and practice. Foundational socio-technical frameworks introduced in the early weeks, such as contextual integrity, behavioral models of decision-making, and privacy-by-design principles, are not treated as isolated concepts. Instead, they are recursively revisited in case analyses, role-play debates, research critiques, and project design decisions. This iterative reinforcement ensures that students continuously apply theoretical constructs to evolving scenarios rather than compartmentalizing knowledge by topic. The structure of the course deliberately integrates multiple modes of engagement. Each lecture establishes analytical vocabulary and conceptual grounding; the case studies require students to operationalize those constructs within realistic institutional and human constraints; and the multi-phase research project demands methodological rigor and synthesis across domains. Guest lectures further strengthen this coherence by exposing students to active researchers and practitioners, situating classroom learning within ongoing scholarly and industry conversations. This integration reinforces authenticity and demonstrates how socio-technical reasoning extends beyond academic abstraction in theory into real-world systems and policy contexts in practice. The coherence of this instructional model has extended beyond the classroom. Student projects and collaborative work emerging from the course have resulted in peer-reviewed publications (redacted for blind review), illustrating that the pedagogical design not only supports learning outcomes but also fosters research contributions of scholarly value. By embedding research production within the learning arc, the course bridges education and knowledge generation. Through this tightly integrated structure, the curriculum cultivates durable socio-technical reasoning rather than fragmented knowledge acquisition. Students leave not only with familiarity of privacy concepts, but with the capacity to critically analyze trade-offs, anticipate unintended consequences, design empirically grounded interventions, and engage ethically with complex technological ecosystems. In doing so, the course positions privacy as an evolving design and research challenge that demands contextual awareness, empirical rigor, and sustained critical inquiry.

\section{Evaluation and Reflections}
To report on the course evaluation, we provide evidence based on end-of-course teaching evaluations including average scores received and a brief summary of the feedback received, covering the students' suggestions. The main ratings are illustrated in section 5.3 in the bar charts showing the scores received in each cohort of the course. Additionally, we provide two instructors' reflections on course development and delivery, and two students' testimonials about their experience and learning outcomes. We invited the instructors to shared their main motivations and thoughts on the course, and the students to briefly explain what they learned and liked in the process.

\subsection{Instructors' Testimonials}

\textbf{Instructor 1} developed the course curriculum grounded in the book~\textit{Modern Socio-Technical Perspectives on Privacy}~\cite{knijnenburg2022modern}. The text offers a comprehensive and interdisciplinary examination of how privacy is conceptualized and operationalized within contemporary digital societies shaped by telecommunications, mobile devices, and pervasive data infrastructures. The course was designed to address students' need for rigorous training in usable privacy by combining theoretical foundations with the cultivation of critical reasoning skills. Rather than limiting instruction to conceptual frameworks, the curriculum encouraged students to interrogate emerging privacy challenges and propose innovative, research-informed responses. Key topics included modeling user intentions, threat modeling, privacy nudging, privacy calculus, and privacy-by-design principles, each examined alongside their underlying motivations, assumptions, and limitations. The integration of guest lectures, structured case studies, and role-playing exercises proved essential in equipping students with practical analytical tools. These components enabled students to approach privacy not merely as a technical safeguard, but as a complex socio-technical challenge requiring pragmatic and proactive problem framing that accounts for user needs, societal implications, and technological constraints. The inclusion of guest speakers from both academia and industry further aligned with the course objectives by bridging scientific inquiry with real-world implementation across domains.

\textbf{Instructor 2} taught the second iteration of the course (the second cohort), building upon the original structure while introducing refinements informed by student feedback and classroom observations. While preserving the socio-technical foundation and emphasis on active learning, Instructor 2 strengthened the alignment among learning objectives, instructional activities, and assessment strategies. The second offering expanded engagement with AI-mediated systems, large-scale data inference, and contemporary platform ecosystems to support deeper exploration of emerging privacy risks. Additional real-world case examples and structured reflection prompts were incorporated to help students more explicitly connect foundational theories to rapidly evolving technological contexts. These refinements resulted in more focused class discussions and more nuanced articulation of stakeholder trade-offs. Structured role plays and case-based debates remained central to the pedagogy, and students demonstrated increased sophistication in analyzing tensions among usability, regulation, accessibility, institutional incentives, and ethical responsibility. Enhanced milestone structuring within the multi-phase research project further reinforced methodological rigor while preserving space for creativity and independent inquiry. Overall, the second iteration demonstrated the adaptability, resilience of the course model and underscored the value of iterative pedagogical refinement in sustaining a dynamic, research-informed privacy curriculum. These refinements are actively shaping the Spring $2026$ offering and will serve as a foundation for continued iterative development.

\subsection{Students' Testimonials}
~\textbf{Student 1} enrolled in Cohort 1 (2024) highlighted the value of integrating contemporary real-world privacy issues through structured case studies. They appreciated how role-playing exercises encouraged students to examine these issues from multiple perspectives, which deepened their understanding of the motivations and constraints shaping the decisions of communities, organizations, and institutions. The inclusion of guest speakers from both academia and industry further enriched the learning experience by connecting theoretical concepts to real-world applications and professional practice. They also noted that the final project provided an opportunity to explore a privacy-focused topic of personal interest in depth, enabling them to apply course concepts in a meaningful and sustained way. Overall, they emphasized that the course fostered active intellectual exchange among students, the instructor, and guest speakers, making each session engaging.

Similarly,~\textbf{Student 2} from Cohort 2 (2025) valued the course's emphasis on critical thinking and its deliberate integration of foundational privacy theories with contemporary technological developments. Through structured discussions, interactive exercises, and scenario-based analyses, they were able to examine privacy challenges from multiple stakeholder perspectives, including policymakers, designers, organizations, and end users. This multidimensional approach strengthened their ability to reason through complex trade-offs that characterize real-world privacy decision-making. They also appreciated the opportunity to engage with guest speakers from academia and industry, whose experiences provided insight into how privacy principles are implemented, negotiated, and sometimes constrained in practice. The quizzes and final examination reinforced their conceptual understanding by offering consistent opportunities to revisit and apply core principles. The multi-phase final project allowed them to pursue a privacy-related topic in depth, conduct an independent literature review, perform experimental analysis, and design a pilot user study while reflecting on broader societal implications. Overall, they described the course as fostering an environment of active dialogue and sustained intellectual engagement, making each session insightful.

\subsection{Teaching Evaluation Reports}
In this section, we report on the results of the university-administered teaching evaluation, a standardized and validated questionnaire distributed at the conclusion of each semester for all courses. The instrument contains a core set of university-wide questions that are applied consistently across course offerings and cohorts. These common items assess multiple dimensions of the teaching and learning experience on a 5-point Likert scale, including students' understanding of key concepts, the effectiveness of instructional methods, opportunities for engagement and participation, the quality and timeliness of feedback, clarity of communication, course organization, and the overall classroom environment. The evaluation also includes measures of student self-reported participation and completion of assigned activities, alongside open-ended prompts inviting students to identify valuable aspects of the course, suggest improvements, and note any barriers to learning. In some semesters, instructors may include a limited number of additional questions tailored to specific course objectives; however, the primary dimensions reported here derive from the shared institutional instrument and therefore allow consistent interpretation across cohorts. The authors consulted with the Institutional Review Board (IRB) of George Mason University
and received approval (Study Number: STUDY00001161) to report these findings, as responses are anonymized and presented only in aggregated form without personally identifiable information. 

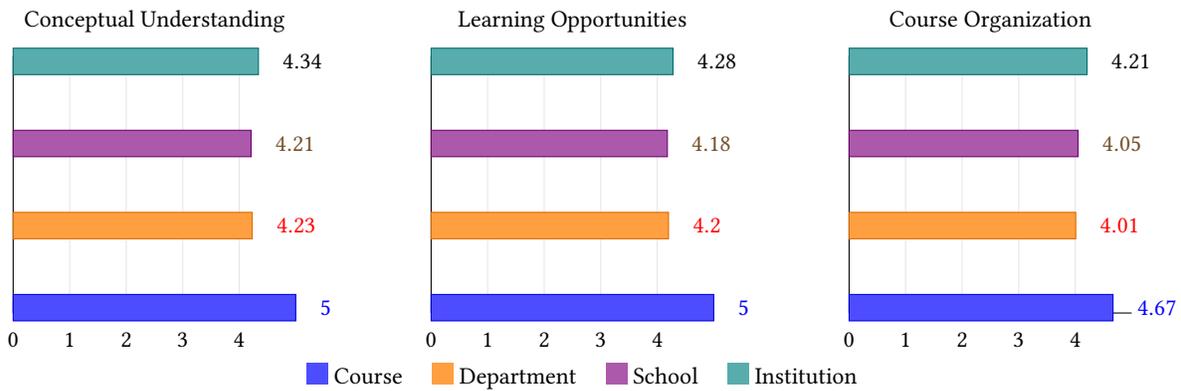
\begin{figure*}[t]
\centering
\begin{tikzpicture}

\begin{groupplot}[
    group style={group size=3 by 1, horizontal sep=1.8cm},
    xbar,
    width=0.3\textwidth,
    height=5cm,
    xmin=0, xmax=5,
    symbolic y coords={Course,Dept,School,Inst},
    ytick=data,
    yticklabels={},                  
    ytick style={draw=none},         
    xmajorgrids,
    grid style={draw=gray!20},
    axis lines*=left,
    xtick={0,1,2,3,4},
    tick label style={},
    xticklabel style={yshift=-4pt},  
    enlarge y limits=0.35,
    nodes near coords,
    nodes near coords align={horizontal},
    every node near coord/.append style={xshift=6pt},
    clip=false
]

\nextgroupplot[title={Conceptual Understanding}]
\addplot+[fill=blue!70,   draw=blue!85!black]   coordinates {(5.00,Course)};
\addplot+[fill=orange!75, draw=orange!85!black] coordinates {(4.23,Dept)};
\addplot+[fill=violet!65, draw=violet!85!black] coordinates {(4.21,School)};
\addplot+[fill=teal!65,   draw=teal!85!black]   coordinates {(4.34,Inst)};

\nextgroupplot[title={Learning Opportunities}]
\addplot+[fill=blue!70,   draw=blue!85!black]   coordinates {(5.00,Course)};
\addplot+[fill=orange!75, draw=orange!85!black] coordinates {(4.20,Dept)};
\addplot+[fill=violet!65, draw=violet!85!black] coordinates {(4.18,School)};
\addplot+[fill=teal!65,   draw=teal!85!black]   coordinates {(4.28,Inst)};

\nextgroupplot[title={Course Organization}]
\addplot+[fill=blue!70,   draw=blue!85!black]   coordinates {(4.67,Course)};
\addplot+[fill=orange!75, draw=orange!85!black] coordinates {(4.01,Dept)};
\addplot+[fill=violet!65, draw=violet!85!black] coordinates {(4.05,School)};
\addplot+[fill=teal!65,   draw=teal!85!black]   coordinates {(4.21,Inst)};

\end{groupplot}

\end{tikzpicture}

{
\textcolor{blue!70}{\rule{8pt}{8pt}} Course \quad
\textcolor{orange!75}{\rule{8pt}{8pt}} Department \quad
\textcolor{violet!65}{\rule{8pt}{8pt}} School \quad
\textcolor{teal!65}{\rule{8pt}{8pt}} Institution
}

\caption{First iteration (Cohort 1) teaching evaluation results (Spring 2024), aligned across three learning dimensions.}
\label{fig:first_iteration_eval}
\Description{Three side-by-side horizontal bar charts labeled Conceptual Understanding, Learning Opportunities, and Course Organization.
Each chart compares course-level ratings with department, school, and institutional benchmarks on a five-point scale.
For Conceptual Understanding and Learning Opportunities, the Course rating is 5.00 and exceeds the benchmarks (approximately 4.18--4.34).
For Course Organization, the Course rating is 4.67, exceeding the department, school, and institutional benchmarks (approximately 4.01--4.21).
Bars are color-coded for Course, Department, School, and Institution and annotated with numeric values.}
\end{figure*}

The purpose of reporting these is to document structured, institutionally grounded evidence of instructional effectiveness and student engagement within a graduate elective context. These data inform iterative course refinement and provide insight into how the pedagogical model performed across consecutive offerings. To report the results of the evaluation forms, the research team plotted the scores regarding conceptual understanding, learning opportunities and course organization for both cohorts (as illustrated in Figures~\ref{fig:first_iteration_eval} and~\ref{fig:second_iteration_eval}). 

\vspace{-2.2mm}

\subsubsection{Teaching Evaluation: First Cohort (Spring 2024)}
In the first offering of \textit{Usable Privacy in the Digital Age} (Spring $2024$), 75\% of the enrolled students completed the university teaching evaluation. The cohort size and response was consistent with an elective, graduate-level special-topics format, responses indicate strong engagement and perceived learning gains. Students reported unanimous top ratings ($5.00/5.00$). These included timely completion of coursework, active contribution to in-class activities, and sustained engagement facilitated by the instructional structure. Students also reported strong learning outcomes tied to the course design and indicated that the course provided meaningful opportunities to acquire key concepts and apply them through discussion-based and activity-driven assignments.

Instructional clarity and support were also rated highly as shown in Figure~\ref{fig:first_iteration_eval}. Students rated clarity of content presentation and opportunities to communicate with the instructor outside class at $4.83/5.00$, and they rated opportunities to provide feedback during the semester at $4.80/5.00$. Ratings for course organization, clarity of expectations, teaching methods, and the helpfulness of feedback averaged $4.67/5.00$, indicating strong perceived structure with room for incremental refinement. Figure~\ref{fig:first_iteration_eval} summarizes the cohort's ratings across three course-level dimensions aligned with the reporting structure used in the second offering. Qualitative responses emphasized the value of guest speakers, case-based discussion, and open-format dialogue. Students recommended expanding explicit coverage of privacy policy and legislation and strengthening links to current events, while reporting no major obstacles.

\subsubsection{Teaching Evaluation: Second Cohort (Spring 2025)}
The second offering of \textit{Usable Privacy in the Digital Age} (Spring 2025) received a 100\% response rate, providing complete evaluation data for the cohort. All respondents were doctoral students enrolled in a face-to-face elective format, ensuring consistent contextual interpretation of results. Across key learning dimensions, the course achieved a mean rating of $4.67/5.00$ for understanding of core concepts, learning through diverse instructional activities, feedback effectiveness, teaching methods, classroom environment, and course organization. Notably, student participation indicators, including completion of assigned tasks and contribution to class discussions received a mean of $5.00$, suggesting strong engagement and ownership of the learning process.

\begin{figure*}[t]
\centering
\begin{tikzpicture}

\begin{groupplot}[
    group style={group size=3 by 1, horizontal sep=1.8cm},
    xbar,
    width=0.3\textwidth,
    height=5cm,
    xmin=0, xmax=5,
    symbolic y coords={Course,Dept,School,Inst},
    ytick=data,
    yticklabels={}, 
    xmajorgrids,
    grid style={draw=gray!20},
    axis lines*=left,
    xtick={0,1,2,3,4},
    tick label style={},
    yticklabel style={},
    xticklabel style={yshift=-4pt},
    enlarge y limits=0.35,
    nodes near coords,
    nodes near coords align={horizontal},
    every node near coord/.append style={xshift=6pt},
    clip=false
]

\nextgroupplot[title={Conceptual Understanding}]
\addplot+[fill=blue!70, draw=blue!85!black] coordinates {(4.67,Course)};
\addplot+[fill=orange!75, draw=orange!85!black] coordinates {(4.26,Dept)};
\addplot+[fill=violet!65, draw=violet!85!black] coordinates {(4.24,School)};
\addplot+[fill=teal!65, draw=teal!85!black] coordinates {(4.35,Inst)};

\nextgroupplot[title={Learning Opportunities}]
\addplot+[fill=blue!70, draw=blue!85!black] coordinates {(4.67,Course)};
\addplot+[fill=orange!75, draw=orange!85!black] coordinates {(4.23,Dept)};
\addplot+[fill=violet!65, draw=violet!85!black] coordinates {(4.21,School)};
\addplot+[fill=teal!65, draw=teal!85!black] coordinates {(4.29,Inst)};

\nextgroupplot[title={Course Organization}]
\addplot+[fill=blue!70, draw=blue!85!black] coordinates {(4.67,Course)};
\addplot+[fill=orange!75, draw=orange!85!black] coordinates {(4.05,Dept)};
\addplot+[fill=violet!65, draw=violet!85!black] coordinates {(4.09,School)};
\addplot+[fill=teal!65, draw=teal!85!black] coordinates {(4.21,Inst)};

\end{groupplot}

\end{tikzpicture}

{
\textcolor{blue!70}{\rule{8pt}{8pt}} Course \quad
\textcolor{orange!75}{\rule{8pt}{8pt}} Department \quad
\textcolor{violet!65}{\rule{8pt}{8pt}} School \quad
\textcolor{teal!65}{\rule{8pt}{8pt}} Institution
}

\caption{Second iteration (Cohort 2, Spring 2025) teaching evaluation results (Spring 2025), aligned across three learning dimensions.}
\label{fig:second_iteration_eval}
\Description{Three side-by-side horizontal bar charts labeled Conceptual Understanding, Learning Opportunities, and Course Organization. 
Each chart compares four benchmark ratings on a five-point scale: Course, Department, School, and Institution. 
In all three dimensions, the Course rating (4.67) exceeds the Department, School, and Institution benchmarks, which range approximately from 4.05 to 4.35. 
The bars are color-coded to distinguish the four comparison groups.}
\end{figure*}
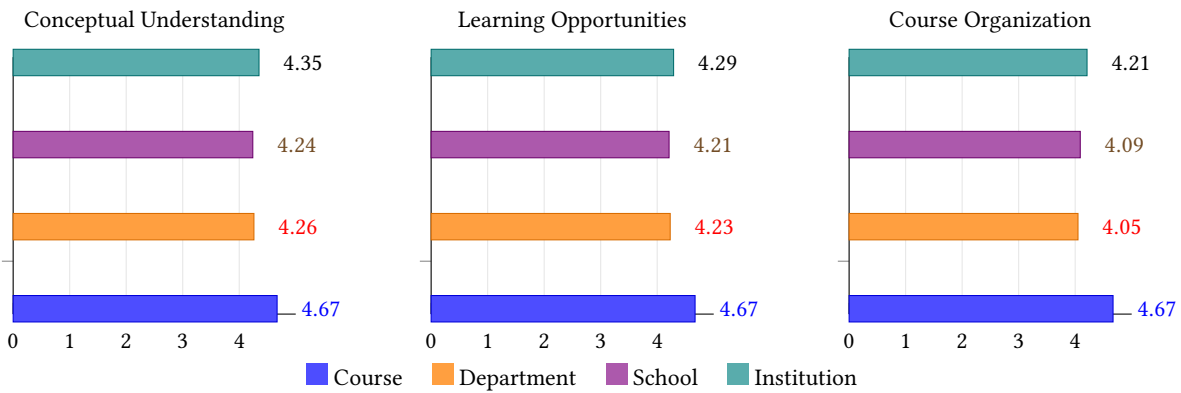

As shown in Figure~\ref{fig:second_iteration_eval}, course-level ratings consistently exceeded department, school, and institutional benchmarks across conceptual understanding, learning opportunities, and course organization. While clarity of communication and presentation received slightly lower means ($4.33$), these scores remain above institutional averages and reflect actionable refinement opportunities rather than structural concerns. Qualitative comments further reinforce these findings. Students identified in-class activities, seminar presentations, guest scholar talks, case studies, and open discussions as the most valuable components of the course. No substantive pedagogical obstacles were reported beyond minor classroom infrastructure limitations. The recommended suggestions for improvement focused on pacing adjustments, including slightly fewer in-class activities and more flexible weekly deliverables. 

Overall, the results of the teaching evaluations show that students were particularly  satisfied with the course contents and structure, highly rating the opportunities for engagement, collaboration, and connection between theoretical and practical aspects of privacy. Additionally, the reflections of the testimonials indicate that there is an opportunity to employ the formative insights to inform and implement incremental refinements in future course offerings.

\section{Future Directions}
In our two iterations of designing, delivering, and evaluating~\textit{Usable Privacy in the Digital Age} suggests several directions for strengthening and scaling privacy education in HCI. While the current course structure successfully integrates socio-technical foundations with active learning and applied research practice, the pace of technological change; particularly around Generative AI, ubiquitous sensing, and platform governance, demands continued evolution in both content and pedagogy~\cite{schlosser2022choosing,ganguly2025generative}. Drawing on reflections from two course offerings, student feedback, and broader trends in research, we outline five strategic directions for future development.

\subsection{Embedding AI-Centered Privacy}
AI systems mediate privacy-relevant interactions by collecting, inferring, and acting upon personal data in ways that are often opaque to users~\cite{postel2024data}. Rather than treating AI as a standalone topic, future iterations of the course will embed AI-centered privacy as a longitudinal theme throughout the semester. This includes sustained engagement with inference risks, secondary data use, model training data governance, and usability challenges in communicating algorithmic decision-making and uncertainty~\cite{afnan2026we}. We will expand case studies and projects to address generative AI, recommender systems, AI assistants in domestic and workplace settings, and privacy implications of automated personalization~\cite{jeckmans2012privacy}. By revisiting AI-related privacy trade-offs across multiple contexts, students will develop a nuanced understanding of how utility, trust, autonomy, and fairness intersect with privacy in intelligent systems.

\subsection{Privacy Engineering and Research-to-Practice Pathways}
Although the course  includes a multi-phase research project, students consistently expressed interest in more hands-on design and evaluation experiences. We therefore plan to introduce a structured ``build-and-evaluate'' module in which students prototype a privacy-enhancing intervention and assess it empirically~\cite{khan2024teaching}. Possible tracks include interface redesign (e.g., consent flows or dashboards), lightweight privacy threat modeling, or integration and evaluation of PETs using user-centered methods~\cite{aly2024tailoring}. In parallel, we will strengthen explicit research-to-practice pathways. Future assignments may include policy briefs, mock privacy impact assessments, or practitioner-facing design memos that require translating research insights into actionable recommendations~\cite{acar2016you}. We also intend to provide additional mentoring for students interested in extending course projects toward conference submissions, thereby reinforcing privacy research as a professional trajectory not only in scientific domains and higher education, but also in industry, standardization or governmental organizations.

\subsection{Enhancing Assessment and Application}
Privacy competence in human-centered education extends beyond conceptual recall; it involves ethical reasoning, stakeholder perspective-taking, and the ability to reason under uncertainty~\cite{robol2018modeling}. To better capture these competencies, future offerings will refine assessment strategies. We plan to develop analytic rubrics for role play and debates that explicitly evaluate argumentation quality, trade-off articulation, and evidence integration~\cite{lazou2025rubric}. Structured reflection prompts will encourage students to revisit earlier positions in light of new evidence or guest lectures, supporting metacognitive development. Additionally, peer- and self-assessment mechanisms will be incorporated into group projects to promote accountability and collaborative reasoning.

Additionally, we aim to facilitate broader adoption of usable privacy education. To this end, we will package and disseminate modular teaching artifacts, including lecture outlines, structured case-study prompts, role-play scenarios, and grading rubrics~\cite{walker2023compulsory,ganguly2026conversational}. ``Plug-and-play'' modules will allow instructors to integrate privacy-focused activities into existing HCI, UX, or security courses without redesigning an entire syllabus~\cite{krawiecka2021plug}. We also aim to curate a living repository of contemporary privacy cases, updated to reflect emerging technologies and regulatory developments~\cite{brodie2005usable}. Such a repository would support both in-person and online delivery formats, providing adaptable materials for synchronous debates, asynchronous discussion forums, and hybrid learning environments.

\subsection{Centering Equity, Global Perspectives, and Diverse Learning Contexts}
Future iterations will further foreground equity-centered and cross-cultural perspectives on privacy~\cite{pujol2021equity,motti2016towards}. This includes a deeper engagement with surveillance disparities, accessibility considerations, and structural power imbalances that shape privacy harms and protections~\cite{lenhart2023you}. Expanding global regulatory and cultural perspectives will help students appreciate how privacy norms and expectations vary across contexts~\cite{agrawal2021exploring}. In addition, we will explore adaptations of the course for different audiences and formats, including advanced undergraduate cohorts, professional short courses, and hybrid or fully online delivery models~\cite{o2008privacy}. Maintaining interactive depth in these formats will require intentional design choices, such as structured breakout debates and asynchronous role-play artifacts~\cite{lim2022mine}. Finally, as part of continuous improvement, we plan to conduct dedicated focus group sessions with instructors and students to systematically gather qualitative insights about course impact, challenges, and areas for enhancement. These reflections will inform iterative revisions and contribute to a more evidence-based model for privacy education in HCI.

\noindent Collectively, these directions aim to strengthen AI-centered content, expand experiential learning, refine assessment of higher-order competencies, scale reusable materials, and foreground equity and global perspectives. Through iterative refinement and community sharing, we seek to advance usable privacy as a foundational pillar of HCI education.

\section{Limitations}
Through this work, we report on the design, implementation, and iterative refinement of a graduate-level usable privacy course, and its evaluation carries several limitations that shape how the findings should be interpreted. First, our empirical evidence is drawn primarily from internal course evaluation mechanisms (standardized end-of-semester teaching evaluations) and qualitative reflections collected within the instructional setting which are limited to two instructors and two students. Although these instruments offer valuable insight into students' perceived learning, engagement, and satisfaction, they do not constitute a robust external assessment of learning gains or skill acquisition beyond the course environment. Secondly, the course enrollments were small across both offerings, which limits statistical power and constrains generalizability. This is typical for specialized graduate electives, but it means the reported patterns should be understood as suggestive rather than definitive. Third, the evaluation emphasizes short-term outcomes observed during or immediately after the course; we did not conduct longitudinal follow-up to assess retention of socio-technical privacy reasoning, transfer of skills to other contexts, or longer-term influence on students' professional practice and research trajectories. Finally, the course was delivered in a specific institutional context (e.g., in-person format, graduate student population, and instructor expertise), which may shape how readily the model transfers to other settings such as larger classes, different student demographics, or online/hybrid delivery.

These limitations do not diminish the value of documenting a practice-oriented, research-informed course model, but they clarify the scope of our claims: we position the contribution primarily as an empirically grounded report of curriculum design and implementation experience, supported by internal evaluation evidence. In future evaluations, we plan to strengthen the evidence base through external pre/post assessments, independent rubric-based evaluation of student artifacts, comparative studies across institutions, focus groups led by external evaluators, and longitudinal follow-ups that measure retention and real-world application over time.

\section{Conclusion}
As data-driven systems normalize pervasive data collection and algorithmic inference, educators must equip students with the knowledge and skills to design, evaluate, and critique privacy-sensitive technologies in realistic socio-technical contexts. In this paper, we presented the design, implementation, and evaluation of~\textit{Usable Privacy in the Digital Age}, a 15-week graduate course that addresses fragmented and theory-heavy approaches to privacy education through an active, practice-oriented pedagogy. We grounded the course in situated learning and integrated structured role playing, case-based discussions, guided forum engagement, guest lectures with reading reflections, and a scaffolded multi-phase research project. Together, these components supported students in connecting foundational privacy theory with real-world decision making. Across two consecutive offerings, our mixed-method evaluation demonstrates strong student engagement and growth in the ability to articulate privacy trade-offs across multiple stakeholders, reason about socio-technical constraints, and apply research methods to emerging privacy challenges. Role-play debates fostered perspective-taking and critical reasoning, while the multi-phase project enabled students to move from literature synthesis to empirical analysis and pilot user study design. This work contributes an empirically informed and reusable model for teaching usable privacy as part of teachable moments. By sharing assignment descriptions, project structures, and grading rubrics, we aim to support replication and adaptation in diverse institutional contexts. As privacy risks evolve alongside AI and ubiquitous technologies, education must remain dynamic, contextual, and practice-driven. Our course offers one step toward building that foundation.


\section{Acknowledgment}
We would like to thank all of the students of this course for their participation and feedback. We would also like to thank the authors of the book - \textit{Modern Socio-Technical Perspectives on Privacy (2022)} - Bart P. Knijnenburg, Xinru Page, Pamela Wisniewski, Heather Richter Lipford,
Nicholas Proferes, and Jennifer Romano. We also thank the guest speakers - Brett Restrick, Yasodara Cordova, Drs. Nathan Malkin, Nata Barbosa, Priya Kumar, Michael Zimmer, Jessica Vitak, Zikai Alex Wen, Filipo Sharevski, Nalin Arachchilage, Renkai Ma, Alisa Frik, and Oshrat Ayalon. This work is partially funded by NSF Grant\#2335636 and \#1937950 and USDA/NIFA Award\#2021-67021-35329. Any opinions, findings, conclusions, or recommendations expressed in this material are those of the authors and do not necessarily reflect the views of the funding agencies. We thank our study participants for their time.

\bibliographystyle{ACM-Reference-Format}
\bibliography{references}

\input{Appendix}

\end{document}

%% file: Appendix.tex
\appendix

\begin{table*}
\centering
\caption{Course Schedule for Spring 2024 and Spring 2025}
\resizebox{\linewidth}{!}{%
\begin{tabular}{|c|l|l|l|l|} 
\hline
Week~ & \multicolumn{2}{c|}{Spring 2024 (cohort 1)} & \multicolumn{2}{c|}{Spring 2025 (cohort 2)} \\ 
\hline
 & \multicolumn{1}{c|}{Topic} & \multicolumn{1}{c|}{Deliverables} & \multicolumn{1}{c|}{Topic} & \multicolumn{1}{c|}{Deliverables} \\ 
\hline
1 & Introduction to Usable Privacy & Assigned Readings & Introduction to Usable Privacy & \begin{tabular}[c]{@{}l@{}}Assigned Readings\\Quiz 1\\Case Study 1\end{tabular} \\ 
\hline
2 & Defining Privacy & \begin{tabular}[c]{@{}l@{}}Assigned Readings\\Case Study 1\\Guest Lecturer 1\end{tabular} & Defining Privacy & \begin{tabular}[c]{@{}l@{}}Assigned Readings\\Case Study 2\end{tabular} \\ 
\hline
3 & Privacy Enhancing Technologies & \begin{tabular}[c]{@{}l@{}}Assigned Readings\\Case Study 2\\Quiz 1\\Guest Lecturer 2\end{tabular} & Privacy Enhancing Technologies & \begin{tabular}[c]{@{}l@{}}Assigned Readings\\Quiz 2\\Case Study 3\end{tabular} \\ 
\hline
4 & Social Media and Privacy & \begin{tabular}[c]{@{}l@{}}Assigned Readings\\Project Discussion\\Case Study 3\end{tabular} & Social Media and Privacy & \begin{tabular}[c]{@{}l@{}}Assigned Readings\\Case Study 4\end{tabular} \\ 
\hline
5 & Tracking, Personalization, and Health & \begin{tabular}[c]{@{}l@{}}Assigned Readings\\Guest Lecturer 3\\Case Study 4\end{tabular} & Tracking, Personalization, and Health & \begin{tabular}[c]{@{}l@{}}Assigned Readings\\Quiz 3\\Case Study 5\end{tabular} \\ 
\hline
6 & Privacy and Vulnerable Populations & \begin{tabular}[c]{@{}l@{}}Assigned Readings\\Guest Lecturer 4\\Case Study 5\end{tabular} & Privacy and Vulnerable Populations & \begin{tabular}[c]{@{}l@{}}\ Assigned Readings\\\ Project - Part 1\\Case Study 6\end{tabular} \\ 
\hline
7 & Accessible Privacy & \begin{tabular}[c]{@{}l@{}}Assigned Readings\\Case Study 6\\Project Phase 1\end{tabular} & Accessible Privacy & \begin{tabular}[c]{@{}l@{}}Assigned Readings\\Quiz 4\end{tabular} \\ 
\hline
8 & Cross-Cultural Privacy & \begin{tabular}[c]{@{}l@{}}Assigned Readings\\Guest Lecturer 5\end{tabular} & Cross-Cultural Privacy & \begin{tabular}[c]{@{}l@{}}Assigned Readings\\Quiz 5 Case Study 7\end{tabular} \\ 
\hline
9 & User-Tailored Privacy & \begin{tabular}[c]{@{}l@{}}Assigned Readings\\Case Study 7\end{tabular} & User-Tailored Privacy & \begin{tabular}[c]{@{}l@{}}Assigned Readings\\Guest Lecturer 1\\Presentation 1\\Case Study 8\end{tabular} \\ 
\hline
10 & The Ethics of Privacy & \begin{tabular}[c]{@{}l@{}}Assigned Reading\\Quiz 3\\Case Study 8\\Guest Lecturer 6\end{tabular} & The Ethics of Privacy & \begin{tabular}[c]{@{}l@{}}Assigned Readings\\Guest Lecturer 2\\Guest Lecturer 3\\Presentation 2\\Quiz 6\\Project – Part 2\\Case Study 9\end{tabular} \\ 
\hline
11 & Bringing Privacy to Practice & \begin{tabular}[c]{@{}l@{}}Assigned Readings\\Case Study 9\\Guest Lecturer 7\end{tabular} & Bringing Privacy to Practice & \begin{tabular}[c]{@{}l@{}}Assigned Readings\\Guest Lecturer 4\\Presentation 3\\Case Study 10\end{tabular} \\ 
\hline
12 & IoT and Wearable Privacy & \begin{tabular}[c]{@{}l@{}}Assigned Readings\\Case Study 10\end{tabular} & IoT and Wearable Privacy & \begin{tabular}[c]{@{}l@{}}Assigned Readings\\Guest Lecturer 5\\Presentation 4\\Quiz 7\\Case Study 11\end{tabular} \\ 
\hline
13 & Smart Homes and Privacy & \begin{tabular}[c]{@{}l@{}}Assigned Readings\\Quiz 4\\Case Study 11\\Guest Lecturer 8\end{tabular} & Smart Homes and Privacy & \begin{tabular}[c]{@{}l@{}}Assigned Readings\\Guest Lecturer 6\\Presentation 6\end{tabular} \\ 
\hline
14 & Critical Reflections on Privacy & \begin{tabular}[c]{@{}l@{}}Assigned Readings\\Seminar Presentations\end{tabular} & \begin{tabular}[c]{@{}l@{}}Critical Reflections on Privacy\\Project Presentations\end{tabular} & \begin{tabular}[c]{@{}l@{}}Project – Parts 1, 2 and 3\\Project Presentation\end{tabular} \\ 
\hline
15 & Project Presentations & \begin{tabular}[c]{@{}l@{}}Project Part 2 Presentations\\Quiz 5\end{tabular} & Final Examination & 2-hour online examination \\
\hline
\end{tabular}
}

\end{table*}